\documentstyle[12pt]{article}

\begin{document}
\setcounter{page}{1}
\begin{titlepage}
\hfill Preprint YERPHI-1571(8)-2001
\vspace{2cm}
\begin{center}

{\bf Pseudoscalar (Charged) Higgs Boson Production with

 $Z^0$ ( $W^{\pm}$ )-Boson  at Muon Colliders in the Models
 
  with Several Higgs Doublets and Singlets}\\
\vspace{5mm}
{\large R. A.  Alanakyan}\\
\vspace{5mm}
{(C) All Rights Reserved \\}
\vspace{5mm}
{\em Theoretical Physics Department,
Yerevan Physics Institute,
Alikhanian Brothers St. 2,

 Yerevan 375036, Armenia}\\
 {E-mail: alanak@lx2. yerphi. am}\\
\end{center}

\vspace{5mm}
\centerline{{\bf{Abstract}}}

 Pseudoscalar and charged  Higgs bosons production in the processes 
 $ \mu^+ \mu^-\rightarrow P^0_i Z^0$, $\mu^+\mu^- \rightarrow H^{\pm}W^{\mp}$
  within the models with several Higgs doublets and singlets is considered.
\vspace{5mm}

\vfill
\centerline{{\bf{Yerevan Physics Institute}}}

\end{titlepage}

                 {\bf 1. Introduction}

As known at muon colliders \cite{P}   Higgs bosons single production in resonance  is possible \cite{H}\cite{BB}:  
\begin{equation}
\label{A1}
 \mu^+ \mu^-\rightarrow H^0(P^0)
\end{equation} 
with large cross section.				 
	However,  mass of Higgs bosons  is not fixed in the
 theory and, thus, we don't know  which energies are
necessary for  Higgs bosons production in resonance.

In this article we continue consideration of pseudoscalar and charged Higgs bosons 
production	 in association with gauge
bosons   in models with Higgs sector   with several Higgs doublets and singlets 
(an examples of such models models see \cite{HK} \cite{GH} e.g. 
in the context of Minimal Supersymmetric Standard  Model (MSSM) and also in the context
 of general two Higgs doublet model and also for supersymmetric models containing 
 two doublets and one singlet of Higgs bosons) .

Previously in \cite{RA1}, \cite{RA}
 has been considered Higgs bosons production in association
 with photon  \footnote {For standard and supersymmetric  Higgs bosons production in association with photons in 
 $e^+e^-$-collisions  see  \cite{BPR},  \cite{DR} ,  \cite{WY} and references therein,  
 scalar Higgs bosons production in association with
 Z-bosons (ZZH) couplings has been considered in    \cite{EGN}, \cite{ER} (see also references therein) .} 
	 at muon colliders:  			 
	 \begin{equation}
	 \label{A1}
	 \mu^+ \mu^-\rightarrow  H^0(P^0) \gamma 
	 \end{equation} 
by model independent way and also has been 
obtained standard Higgs bosons case and 
enhancement in supersymmetric Higgs bosons case in comparison with
  standard Higgs bosons production in  reaction (2) 
  The process (2) (for standard Higgs bosons
 has been  calculated also in \cite{LT}  \cite{ABDR}
however authors of this papers  do not consider enhancement in theories
 with extended Higgs sector and besides authors of \cite{LT} do not consider
  loop effects which as was shown in \cite{RA1} and \cite{RA} are very important. 
  It must be noted also that our analytical  calculations  are  essentially differ from calculations
   of \cite{LT}.	

In \cite{A}  has been considered  supersymmetric charged and pseudoscalar Higgs bosons
 production at muon colliders in
 association  with $W^{\pm},Z^0$-bosons:   		 
\begin{equation}
\label{A4}
 \mu^+ \mu^-\rightarrow H^0_3 Z^0,
\end{equation}
\begin{equation}
\label{A5}
 \mu^+ \mu^-\rightarrow H^{\pm} W^{\mp}
\end{equation}				 
	This processes are similar to the processes calculated 
	in theories   with $R$-parity violation \cite{F}-\cite{D}: 			 
\begin{equation}
\label{A2}
 l^+_i l^-_j \rightarrow \tilde{\nu}_{kL} Z^0,
\end{equation}
\begin{equation}
\label{A3}
 l^+_i l^-_j \rightarrow \tilde{l}^{\mp}_{kL} W^{\pm}.
\end{equation} 
also considered in   \cite{A} \footnote {Previously the processes 
$e^+ e^-\rightarrow  \tilde{\nu}_{kL} Z^0, \tilde{l}^{\mp}_{kL} W^{\pm}$ 
have been discussed in \cite{DL}. However, our results  \cite{A}  are essentially differ from results 
of the \cite{DL}. For instance, formulas for cross section of the process 
$e^+ e^-\rightarrow  \tilde{\nu}_{kL} Z^0$ in  \cite{DL}
do not contain $a_{L,R}$ in contrast to our result. Feynmann diagrams corresponding
 to the process  $e^+ e^-\rightarrow \tilde{\nu}_{kL} \gamma$ has been also discussed in
 \cite{DL} however its cross section  has not been calculated.Case of different flavors 
 of colliding leptons  in this paper also has not been considered .}				 
				 
In this article we consider by model independent way  pseudoscalar  (charged) Higgs bosons
production at lepton antilepton colliders
 in association with $Z^0(W^{\pm})$ bosons in model which  contain many
 pseudoscalars and many scalars 
(such large number of scalars and pseudoscalars appear e.g.  
in models which Higgs sector consist of  several doublets + several singlets  ).  
Part of the scalars  may have masses so that resonances in $S$ -channel diagram
 is possible( in contrast to MSSM)  .I.e.  condition $m_j >m_P+m_ Z$  for some 
 masses of scalars  $m_j$ are possible.

 From the Fig.2 of the paper \cite{BB} we see that after taking into account radiative corrections
 (for radiative corrections see references
   in \cite{BB})
   situation  $m_P+m_Z<m_H$ within the MSSM is possible.

      Due to this possibility  as was shown in this article   resonant production (1)  
    $\mu^+\mu^-\rightarrow H_1^0$
     with subsequent decay $H_1^0 \rightarrow Z^0+H_3^0$ is possible     ($H_1^0$
    is heaviest scalar Higgs boson in the MSSM, $H_3^0$ 
	is  psudoscalar Higgs boson in the MSSM ).This situation take place at small $tan\beta \sim 1$ 
	and not heavy  psudoscalar  ($m_P<  50 -100 $ GeV depend on scalar quarks masses and $ tan \beta$.
	For example at $ tan \beta \approx 1$, stop mass 1 TeV possible  $m_{H_3}=100$ GeV $m_{H_1}=200$ GeV
	at    stop mass 0.5 TeV   $m_{H_3}<50$ GeV       $  m_{H_1}=140$ GeV ). 
	I.e. studied situation is possible even in MSSM.

Situation which considered in this paper  is similar to the consideration of the analogous process
 $l^+_i l^-_j \rightarrow \tilde{l}^{\mp}_{kL} W^{\pm}$
in which this phenomena (resonance enhancement via intermediate sneutrino   exchange even
 far from resonance 
  \footnote  {for sneutrino 
  resonant production see   \cite {BGH}-  \cite{EF} },
  scalar neutrino is scalar + pseudoscalar with same masses) 
  also take place  \cite{A}.There are many parallels between scalar neutrinos (scalars leptons)
   production in theories with R-parity violation and neutral (charged) Higgs bosons.
   
   \newpage
    
   Also related phenomena
    has been found in   \cite{AB}  where in the framework of
  Two Higgs -doublet Model has been considered enhancement of the  process  (4)  
 ( in comparison with previous  consideration of the process (4)  in the framework of MSSM in \cite{A}  ) 
 which appear due to resonance of intermediate scalar and pseudoscalar Higgs bosons on  
 $S$ -channel diagramm.

{\bf 2. Results for $ \mu^+ \mu^-\rightarrow P^0_i Z^0$}

In our model independent   calculations we suppose that the couplings of the scalar and pseudoscalar
Higgs bosons couplings to leptons are following:
\begin{equation}
\label{1}
{\cal L}=ih_j \bar{l}l+h_P \bar{l } \gamma _5 l  P^0
\end{equation}
Interaction  $Z^0H_j^0P^0$ is  following:
\begin{equation}
\label{A6}
  h_{jZ}(k_4 \times Z) 
\end{equation}

Our results for differential cross section of the process $ \mu^+ \mu^-\rightarrow P^0_i Z^0,$ are 
following:
\begin{equation}
\label{A9}
 \frac{d\sigma( \mu^+ \mu^-\rightarrow P^0_i Z^0) }{dt}=
\frac{\alpha }{8 \sin^2 \theta_W \cos^2\theta_W s^2}
(B_1+B_2+B_3),
\end{equation}
where:
\begin{equation}
\label{A9}
 B_1=\frac{s}{4m_Z^2}|b|^2-m_P^2s| \sum _j \frac{h_jh_{jZ}}{s-m_j^2+i\Gamma_jm_j}|^2,
\end{equation}
\begin{equation}
\label{A9}
 B_2=\frac{sm_P^2}{2}(\frac{1}{t}+\frac{1}{u}) h_P \sum _j  Re(\frac{h_jh_{jZ}}{s-m_j^2+i\Gamma_jm_j}),
\end{equation}

\begin{equation}
\label{A9}
 B_3=h_P^2[(a_L^2+a_R^2)(\frac{1}{t^2}+\frac{1}{u^2})(tu-m_P^2m_Z^2)+
\frac{4a_La_R(t-m_P^2) (u-m_P^2)}{tu}],
\end{equation}

\begin{equation}
\label{A9}
 b=h_P+  (s-m_P^2-m_Z^2)\sum _j  \frac{h_jh_{jZ}}{s-m_j^2+i \Gamma_j m_j}
\end{equation}
Requirement of unitarity lead to the condition:
\begin{equation}
\label{A9}
 0=h_P+\sum _j h_jh_{jZ}
\end{equation}
\begin{equation}
\label{A13}
t_-<t<t_+,
\end{equation}
where
\begin{equation}
\label{A14}
t_{\pm}=\frac{m_P^2+m_{Z}^2-s\pm
\sqrt{(m_{P}^2+m_{Z}^2-s)^2-4m_{P}^2m_{Z}^2}}{2}.
\end{equation}

After performing integration within the limits (13), (14) we obtain for the total
cross sections the following result:
\begin{equation}
\label{A15}
\sigma( \mu^+ \mu^-\rightarrow P^0_i Z^0)=
\frac{\alpha }{4 \sin^2  \theta_W\cos^2 \theta_W s^2}
( A_1\log(\frac{t_+}{t_-})+A_2(t_+-t_-)),
\end{equation}

where:
\begin{equation}
\label{A15}
 A_1=h_P^2(a_L^2+a_R^2)(m^2_P+m^2_Z-s)+2a_La_R \frac{m_P^2 (s-m_Z^2)}{m^2_P+m^2_Z -s}
 +\frac{sm_P^2}{2}\sum _j  h_Ph_jh_{jZ}
\frac{(s-m_j^2)}{(s-m_j^2)^2+\Gamma_j^2m_j^2},
\end{equation}

\begin{equation}
\label{A15}
 A_2=\frac{s}{8m_Z^2}|b|^2-\frac{sm_P^2}{2}|\sum _j   \frac{h_jh_{jZ}}{(s-m_j^2)+i \Gamma_j m_j}|^2+
 2h_P^2(a_La_R-(a_L^2+a_R^2)),
\end{equation}

In the vicinity of resonance of the one of the scalar on the $s$-channel diagram  
($s\approx  m_j^2 $) we obtain  Breit Wigner cross section (as in \cite{BB}) :
\begin{equation}
\label{A15}
\sigma( \mu^+ \mu^-\rightarrow P^0_i Z^0) \approx
\frac{4 \pi \Gamma (H_j^0\rightarrow \mu^+\mu^-)  \Gamma(H_j^0 \rightarrow Z^0 P^0)}
{(s-m_j^2)^2+\Gamma_j^2m_j^2}
\end{equation}
as it must be.

In the limit $s, m_{j}^2 \gg  m_{P}^2$      ($m_{j}$ -is arbitrary)        the previous
 formulas are reduced and we have:

\begin{equation}
\label{A17}
\sigma( \mu^+ \mu^-\rightarrow P^0_i Z^0 )
=\frac{\alpha }{4 \sin^2 \theta_W \cos^2 \theta_W s}
(2 h_P^2   (a^2_L+a^2_R)
s \log(\frac{s}{m_{P} m_Z})+A_2),
\end{equation}
In the MSSM at large $tan \beta$  it is necessary to put $h_P=\frac{gm_{\mu }tan \beta}{\sqrt{2}m_W}$ 
and besides mass of scalar is equal to the mass of pseudoscalar and we obtain formula (25) of 
\cite{RA}   in the limit  $s \gg  m_{j}^2,  m_{P}^2$ .

We will present numerical results for specific models in our next article (for example for two Higgs
 doublet model). However it is obviously 
that  even off resonance  the cross section may be essentially enhanced. 
The situation is very similar to the \cite {RA}(see Fig.4 ) where due to resonance of scalar neutrino exchange
 the process (2,3) essentially enhanced even far from resonance.  Analogous enhancement
   in the process (4) has been  obtained in \cite{AB} in the framework of two HDM model (see Fig.4 in this paper).

{\bf 3. Results for $ \mu^+ \mu^-\rightarrow  H^{\pm} W^{\mp}$}
 
 In our    model independent   calculations we suppose that the couplings of the charged
Higgs bosons  to leptons are following:
\begin{equation}
\label{A6}
{\cal L}=h \bar{l } P_L \nu H^- +h.c. +\sum _j  h_j \bar{l }l+\sum _j  h_{Pj} \bar{l } \gamma _5 l
\end{equation}
Interactions  $W^+H_j^0H^-$ and $W^+P_j^0H^-$   are  following: 
\begin{equation}
\label{A6}
H_{jW} W^+H_j^0H^-,  P_{jW} W^+H_j^0H^-
\end{equation}
Amplitude of the process (3) may be written as :
\begin{equation}
\label{A22}
M=\frac{g}{\sqrt{2}} \bar{u}(k_1)[( h \frac{\hat{k}_4
\hat{W}}{t}+(k_4W) a_+(j)H_{jW}h_j
) P_{L}+((k_4W)  \sum _j   a_-(j)H_{jW}h_j
) P_{R}]u(k_2)              .
\end{equation}
where:
\begin{equation}
\label{A22}
 a_{\pm}(j)=\frac{1}{s-m_j^2+i\Gamma_jm_j} \pm \frac{1}{s-M_j^2+i\Gamma_j'M_j}
\end{equation}
where $\Gamma _ j, \Gamma _ j'$ are widths of scalars and pseudoscalars ,  $m_j, M_j$ are masses of 
scalars and pseudoscalars.

Requirement of unitarity lead to the conditions:
\begin{equation}
\label{A9}
 0=-h+\sum _j  \frac{1}{2}h_jH_{jW}
\end{equation}
\begin{equation}
\label{A9}
 0=h+ \sum _j  \frac{1}{2}h_{Pj}P_{jW}
\end{equation}
Formulas presented in \cite{AB} for cross section of the process (4) jointly 
with formulas (26),(27) of this paper
 may be used in principle
 for any Higgs sector  although in this paper
  has  not been  mentioned this possibility. In particularly in theories with more than two doublets
   and one or more singlets may exist several resonances of the scalars and
    several resonances of pseudoscalar  which may very essentially to enhance the corosse section of the process (4).

 {\bf 4. Loop Diagrams}
 
 Besides tree diagrams also exist loop diagrams (Fig.2)  which also contribute to the processes
   
$l^+l^- \rightarrow P^0 Z^0$,$ l^+l^- \rightarrow H^{\pm}W^{\mp}$ .

Blocks  $H^+W^-Z^0, Z^0Z^0P^0,H^+W^- \gamma $$H^0Z^0\gamma$,
    $P^0Z^0\gamma$, $H^-\mu^+\nu$, $H^0_i\mu^+\mu^-$ on diagrams on the Fig.2 has been
     considered previously in  
		  (\cite{MP}-\cite{WY}, \cite{BB}(see also refernces therein).
	 For loop vertexes   $H^0Z^0\gamma$,
    $P^0Z^0\gamma$     and full process  
	 process $e^+e^-\rightarrow H^0(P^0)+photon$  see   \cite{WY} and references therein.
	 In papers \cite{BC} and  \cite{HS}  has been considered process 
	  $e^+e^- \rightarrow P^0 Z^0$ due to 
		  block $Z^0 Z^0P^0$, all another diagrams of Fig.2 in this papers has
		   not been considered.
	 
	  In  \cite{A} , (and in  \cite{A2} in the form presented in this article 
  on the Fig.2) has been considered also box diagrams and diagrams
   containing  $H^-\mu^+ \nu$, $H^0_i \mu^+\mu^-$
-vertexes  (including supersymmetric particles contribution: charginos ,neutralinos, scalar neutrinos , 
charged scalar leptons...) 
with some estimates and conclusions  where this diagrams may be important. Obviously this loop diagrams
 are dominant  in $e^+e^-$ annihilation where tree diagrams are nonimportant due to small mass of electron.  
	 
It is interesting to compare  our tree  results and estimates  of loop contribution to the processes
$\mu^+\mu^- \rightarrow P^0 Z^0$, $ \mu^+\mu^- \rightarrow H^{\pm}W^{\mp}$
 of the Fig.2  o
with exact calcultions  of the processes 
 $e^+e^- \rightarrow P^0 Z^0$, $ e^+e^- \rightarrow H^{\pm}W^{\mp}$ 
 also based on diagrams of the Fig.2  which has been performed
 in the later papers (\cite{AAC} ,\cite{AAC1},\cite{K},\cite{Zh},\cite{AAC1}, and  in paper \cite{AAC1} 
 within  MSSM  with SUSY particles in loops).

 For example from paper \cite{Zh} we see that cross section of charged Higgs bosons production+W-bosons 
 via loops in $e^+e^-$-annihilation is less than   0.01 fb  at $\sqrt{s}=500 GeV$ and $tan \beta>10$
  and decrease with $tan\beta$ growth.
 At $\sqrt{s}=1000 GeV$ and $tan \beta>10$  the author of this paper obtain  the cross section  less than 0.003 fb. 
 
   From paper  \cite{K} we see that cross section of the process $e^+e^- \rightarrow H^{\pm} W^{\mp}$
  is smaller than 0.003 fb in the MSSM case at    $tan \beta>10$ .From Fig.5 of this paper we see that cross section 
  is  smaller than  o.1 fb at  $tan \beta >10$    in 2HDM case and essentially decrease with growth  $tan \beta $.

From paper \cite{AAC} we see that number of psedoscalars +$Z^0$-bosons produced via loop of Fig.2
 (without SUSY contribution)  at  $\sqrt{s}=500GeV$ and $10<tan \beta $ 
 the cross section $<0.001 fb$.At  $6<tan \beta \beta$ the cross section $<0.01 fb$ has been obtained.
 
 From paper \cite{AAC1} where has been considered process $e^+e^-\rightarrow Z^0P^0$
  (loop contribution with SUSY particles as our diagram 2)
  we see that again the cross section is smaller than 0.1 fb in most favorable cases  in 2HDM   model.
  In case of MSSM the cross section is smaller than 0,01 fb in most favorable range of parameters.
 However it very difficult to consider all ranges of parameters and probably in some area various SUSY contribution
    will even to cancel to each over  (see discussion in \cite{RA}). The  authors  of   \cite{AAC1}  consider 
	 only several set of parameters and probably  at another parameters the situation will be essentially differ.
	 Such partial cancellation  depend on SUSY parameters take place in the loop process 
	 $e^+e^- \rightarrow  photon +P^0 (H^0)$.
	  
In accordance with 	  \cite{A} 
 total cross sections of processes (3), (4), may be obtained
 from   cross section of the processes of scalar neutrinos and leptons  ( 
 $l^+_i l^-_j \rightarrow \tilde{\nu}_{kL} Z^0,
 l^+_i l^-_j \rightarrow \tilde{l}^{\mp}_{kL} W^{\pm}
 $) by the following replacements (because scalar and pseudoscalar Higgs bosons
  interactions with leptons are similar to  scalar leptons  interaction with leptons )   :
\begin{equation}
\label{A22}
h_{ijk}\rightarrow \frac{g m_{\mu}}{\sqrt{2}m_W}\tan\beta
,  m_{\tilde{\nu}}\rightarrow m_3,
m_{\tilde{l}}\rightarrow m_4 \quad and \quad \Gamma_{\tilde{\nu}}=0
\end{equation}
where  $m_{\mu}$ is $\mu$-meson mass.
To value   $h_{ijk}=10^{-2} $ of Yukawa coplings   of scalar neutrino and lepton to ordinary leptons 
 correspond $tan\beta=17.5$ and we obtain  form Fig.4,5 of \cite{A} 
that  about
 $\approx 1200 (\frac{\tan\beta}{17. 5})^2- 60(\frac{\tan\beta}{17. 5})^2$
events per year at $L=1000$  $fb^{-1}$-$50$  $fb^{-1}$,  $\sqrt{s}=1000$   $GeV$,  
$m_4=300-700$  $GeV$. About 1100 events per year at $L=1000$  $fb^{-1}$-$50$  $fb^{-1}$,  
$\sqrt{s}=500$   $GeV$,  $m_4=300$  $GeV$.
 
For  MSSM pseudoscalar Higgs production with $Z^0$-bosons in  \cite{A}
has been obtained
$\approx 600  (\frac{\tan\beta}{17. 5})^2- 30(\frac{\tan\beta}{17. 5})^2$ pseudoscalar Higgs bosons
per year  at            .$\sqrt{s}=500-1000$   $GeV$ and masses $m_P=500-700$ GeV.At kinematical
  peak we have  about $1000$  events per year  at same $\tan\beta$ and  $m_P=500$ GeV.
  At smaller  $m_P$ the cross section may be also enhanced. 

In any cases we see that at  $tan \beta>10$  ( and sometimes at smaller $tan \beta $ )  
the tree contribution considered in \cite{A} in the processes (3),(4)
  is much  more significant than loop contributions considered in  results of this papers
   \cite{AAC},\cite{AAC1},\cite{K},\cite{Zh}.
    
	 Thus we confirm our previous result \cite{A} in accordance with at $tan \beta \sim 7$ or higher tree
 contribution is more essential than loop contribution.
 
	 However if resonances at  third s-channel  diagrams of the processes
	  (3) is exist  ( i.e.  masses of some of the scalars is larger than $m_P+m_Z$) 
the cross section of the tree process (3) may be essential even at  small Higgs bosons 
interactions with leptons. 

The author express his sincere gratitude to  G. K. Yeghiyan for helpful discussions.

\centerline {\bf Figures captions:}

Fig. 1 Tree  diagrams corresponding to the processes (2), (3).$H^0$ 
 denoted scalars $P^0_j$-pseudoscalars. 

Fig. 2  Loop diagrams corresponding to the processes (2), (3) within the MSSM..
 Shaded ring corresponds  to the diagrams with loop-induced
$\gamma W^{\mp}H^{\pm}$, $Z^0W^{\mp}H^{\pm}$,  
$\gamma Z^0H^0_i$, $Z^0Z^0H^0_i$, $H^-\mu^+\nu$, $H^0_i\mu^+\mu^-$
-vertexes.Diagrams with loop-induced  virtual gauge bosons-Higgs bosons mixing are not shown. 
\end{document}